\documentclass[aps,twocolumn,showpacs]{revtex4}
\usepackage{graphicx}
\usepackage{amsmath}
\usepackage{amsfonts}
\usepackage{color}
\newcommand{\tbox}[1]{\mbox{\tiny #1}}

\begin{document}



\title{Fuzzy Erd\"os-R\'enyi random graphs: Spectral statistics}


\author{J. A. M\'endez-Berm\'udez}
\address{Instituto de F\'isica, Benem\'erita Universidad Aut\'onoma de Puebla, Puebla 72570, Mexico}

\begin{abstract}
Within a random-matrix-theory (RMT) approach, we perform a detailed numerical study of the spectral statistics
of fuzzy Erd\"os-R\'enyi (ER) random graphs.
Specifically, we consider the membership values of the vertices and edges of a fuzzy ER graph model as random 
variables. In this way, the corresponding adjacency matrices become sparse matrices with random entries.
Then, we apply standard RMT measures to characterize the eigenvalues and eigenvectors of ensembles of 
Hermitian and non-Hermitian adjacency matrices of both undirected and directed fuzzy graphs.
We demonstrate that the average degree of the ER model serves as the scaling parameter of the spectral 
properties of the corresponding fuzzy setups.
\end{abstract}

\maketitle

\section{Preliminaries}

Fuzzy graphs generalize the concept of a graph by allowing for the partial membership of the graph components 
(vertices and edges)~\cite{R75,MN00}. This approach is particularly useful in modeling systems with uncertainty 
or when the graph components are simply non equivalent.

Even though, the concept of fuzzy graph has been introduced long ago by A. Rosenfeld~\cite{R75}, most of 
the studies on fuzzy graphs focus on structural properties  (see e.g.~\cite{SA24,SAB19,SM13,SS16,CRKK25}) 
and just a few on the spectral properties of the corresponding adjacency matrices; see e.g.~\cite{RRASA24,SSR24,VJ16,AAA23}.

It is relevant to note that allowing partial memberships of the vertices and edges of a graph produces, 
straightforwardly, weighted adjacency matrices; see e.g.~\cite{RRASA24,CRKK25,SSR24,VJ16,AAA23}. 
Moreover, if those membership values are random numbers
drawn from a given probability distribution, the corresponding randomly-weighted adjacency matrices may
be considered as members of random matrix ensembles similar to the standard ensembles of random matrix 
theory (RMT). Indeed, in this work, we consider the membership values of the vertices and edges of a fuzzy 
Erd\"os-R\'enyi graph model as random variables, see Sec.~\ref{model}. In this way, the corresponding 
adjacency matrices become sparse matrices with random entries. Then, within a statistical RMT approach, we 
apply standard RMT measures
(defined in Sec.~\ref{measures}) to characterize the eigenvalues and eigenvectors of ensembles of Hermitian 
and non-Hermitian adjacency matrices corresponding, respectively, to undirected and directed graphs, see 
Sec.~\ref{results}.

\section{Model and measures}

\subsection{The fuzzy Erd\"os-R\'enyi random graph model}
\label{model}

An Erd\"os-R\'enyi (ER) random graph $G_{\tbox{ER}}(n,p)=(V_{\tbox{ER}},E_{\tbox{ER}})$ consists of 
$n$ vertices $v_i\in V_{\tbox{ER}}$ where each edge $e_{ij}\in E_{\tbox{ER}}$ connecting the pair of 
vertices $(v_i,v_j)$ appears independently with probability $p \in [0,1]$.
We define a fuzzy ER random graph as follows:
Let $\sigma$ be a fuzzy set on the vertex set $V_{\tbox{ER}}$ and $\mu$ be a fuzzy set on the edge set 
$E_{\tbox{ER}}$.
Here, we consider $\sigma_i(v_i)$ as statistically-independent random variables drawn from a flat distribution 
in the interval $[0,1]$ and $\mu_{i,j}(e_{i,j})$ as statistically-independent random variables drawn from a 
flat distribution in the interval $[0,\min(\sigma_i,\sigma_j)]$.
That is, the membership values of the vertices and edges of this fuzzy ER graph model are random variables.
See also Ref. \cite{KS26} for the definition of fuzzy ER models.

The prescription above naturally leads to a randomly-weighted adjacency matrix:
\begin{equation}
[\mathbf{A}]_{ij}=\left\{
\begin{array}{cl}
\sigma_i & \mbox{if $i=j$}, \\
\mu_{ij} & \mbox{if vertices $v_i$ and $v_j$ are connected}, \\
0 & \mbox{otherwise}.
\end{array}
\right.
\label{A}
\end{equation}
Notice that the random matrix $\mathbf{A}$ has two limits: When $p=0$ it is a diagonal random matrix,
while it is a full random matrix for $p=1$. However, for any $p\in (0,1)$ it is a sparse random matrix.
Then, we expect to observe a transition form a diagonal random matrix to a full random matrix by
increasing $p$ from 0 to 1, for any given $n$.

Moreover, below we consider two setups: Undirected and directed fuzzy ER random graphs.
In the undirected (directed) case the corresponding adjacency matrix is Hermitian (non-Hermitian), 
$\mathbf{A_{\tbox{H}}}$ ($\mathbf{A_{\tbox{nH}}}$).

It is relevant to stress that when $p=1$, the full random matrices $\mathbf{A_{\tbox{H}}}$ and 
$\mathbf{A_{\tbox{nH}}}$ do not correspond to any standard RMT ensemble. 
However, we will use as reference the predictions for the Gaussian Orthogonal Ensemble (GOE) 
(i.e.~full real and symmetric matrices with gaussian random entries) and the Real Ginibre Ensemble 
(GinOE) (i.e.~full real and non-symmetric matrices with gaussian random entries) which are similar,
in the sense of hermiticity, to $\mathbf{A_{\tbox{H}}}$ 
and $\mathbf{A_{\tbox{nH}}}$ when $p=1$, respectively.

\subsection{Random matrix theory measures}
\label{measures}

In this work we characterize the eigenvalue and eigenvector properties of the randomly-weighted adjacency 
matrices of Eq.~(\ref{A}) by the use of well-known RMT measures.

On the one hand we characterize the real spectra of the Hermitian matrix 
$\mathbf{A_{\tbox{H}}}$ and the complex spectra 
of the non-Hermitian matrix $\mathbf{A_{\tbox{nH}}}$ by computing, respectively, the average value 
of the eigenvalue spacing ratio $r_\mathbb{R}$ and the average value of 
the ratio between nearest- and next-to-nearest neighbor eigenvalues $r_\mathbb{C}$, which are 
defined as follows. Given the real ordered spectrum 
$\lambda_1>\lambda_2>\cdots>\lambda_{n-1}>\lambda_n$, the $i$-th ratio $r_\mathbb{R}^i$ 
reads as~\cite{OH07,ABGR13}
\begin{equation}
\label{rR}
r_\mathbb{R}^i = \frac{\min(\lambda_{i+1}- \lambda_i,\lambda_{i}- \lambda_{i-1})}{\max(\lambda_{i+1}- \lambda_i,\lambda_{i}- \lambda_{i-1})} \ .
\end{equation}
Here, $r_\mathbb{R}\in[0,1]$.
Given the complex spectrum $\{ \lambda_i \}$ the $i$-th ratio $r_\mathbb{C}^i$ 
reads as~\cite{SRP20}
\begin{equation}
r_\mathbb{C}^i = \frac{\left| \lambda^{NN}_i - \lambda_i \right|}{\left| \lambda^{NNN}_i - \lambda_i \right|} \ ,
\label{rC}
\end{equation}
where $\lambda^{NN}_i$ and $\lambda^{NNN}_i$ are, respectively, the nearest and the next-to-nearest 
neighbors of $\lambda_i$ in $\mathbb{C}$. Note that, as well as $r_\mathbb{R}$, $r_\mathbb{C}\in[0,1]$. 
In the case of random graphs and networks, $\left\langle r_\mathbb{R} \right\rangle$ and 
$\left\langle r_\mathbb{C} \right\rangle$ have been used as complexity indicators able to characterize the 
transition between isolated vertices and complete graphs; see e.g.~\cite{AMRS20,PRRCM20,PM23}.

On the other hand we characterize the eigenvectors of both $\mathbf{A_{\tbox{H}}}$ and  
$\mathbf{A_{\tbox{nH}}}$ by computing their Shannon entropies $S$ and inverse participation ratios
$\mbox{IPR}$, which are defined as follows.
For the $i$-th normalized eigenvector $\Psi^i$, i.e. $\sum_{j=1}^n | \Psi^i_j |^2 =1$, we have~\cite{S48}
\begin{equation}
\label{S}
S_i = -\sum_{j=1}^n \left| \Psi^ii_j \right|^2 \ln \left| \Psi^i_j \right| ^2 
\end{equation}
and~\cite{OH07} 
\begin{equation}
\label{IPR}
\mbox{IPR}_i = \left[ \sum_{j=1}^n \left| \Psi^i_j \right|^4 \right]^{-1}.
\end{equation}
Both $S$ and $\mathrm{IPR}$ measure the extension of eigenvectors on a given basis.
In the case of random graphs, $\left\langle S \right\rangle$ has been used to prove percolation transitions 
(see e.g.~\cite{MAMRP15,PRRCM20,PM23}), while $\left\langle \mbox{IPR} \right\rangle$ is commonly 
used to characterize the localization properties of a network (see e.g.~\cite{ARM18,FMRM20,GDOM12}).

\section{Numerical results}
\label{results}

In this section we  perform a detailed numerical study of the spectral statistics
of fuzzy ER random graphs. Moreover, we consider undirected and directed graphs separately.

\subsection{Undirected fuzzy Erd\"os-R\'enyi random graphs}

In what follows we use exact numerical diagonalization to obtain the eigenvalues $\lambda_i$ and
eigenvectors $\Psi^i$ ($i =1,\ldots,n$) of ensembles of Hermitian adjacency matrices $\mathbf{A_{\tbox{H}}}$
characterized by the parameter pair $(n,p)$.

Since the spectrum of $\mathbf{A_{\tbox{H}}}$ is real, the natural choice to characterize it is by computing
$\left\langle r_\mathbb{R}(\mathbf{A_{\tbox{H}}}) \right\rangle$. Moreover, note that $r_\mathbb{C}$ can 
also be computed for real spectra, so we complete our analysis by also computing  
$\left\langle r_\mathbb{C}(\mathbf{A_{\tbox{H}}}) \right\rangle$.

\begin{figure}[t]
\centering
\includegraphics[width=0.95\columnwidth]{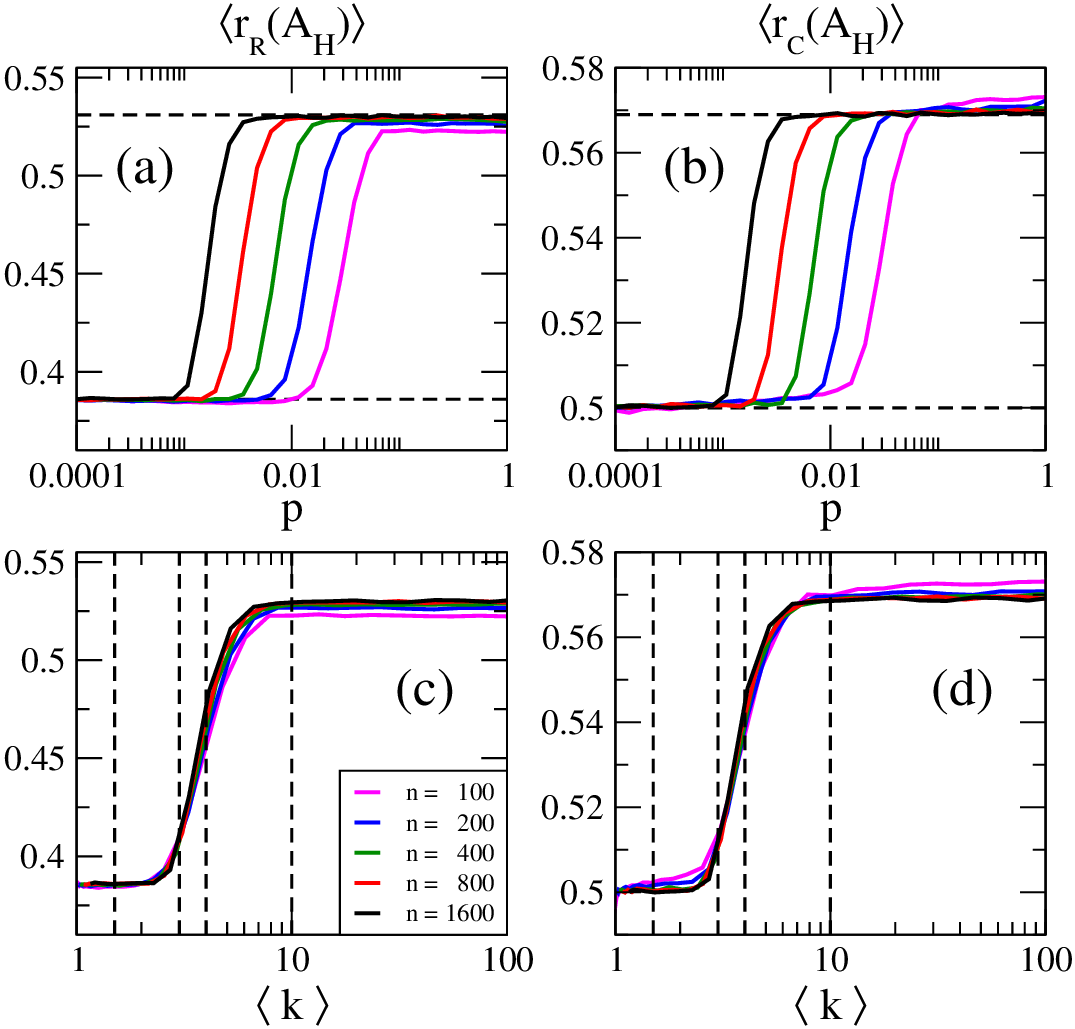}
\caption{Average ratios 
(a,c) $\left\langle r_\mathbb{R}(\mathbf{A_{\tbox{H}}}) \right\rangle$ and 
(b,d) $\left\langle r_\mathbb{C}(\mathbf{A_{\tbox{H}}}) \right\rangle$ of 
undirected fuzzy Erd\"os-R\'enyi random graphs (of size $n$) 
as a function of (a,b) the probability $p$ and (c,d) the average degree $\left\langle k \right\rangle$.
Horizontal dashed lines in (a) correspond to the PE and GOE predictions for 
$\left\langle r_\mathbb{R}(\mathbf{A_{\tbox{H}}}) \right\rangle$:
$\langle r_{\mathbb{R}}\rangle_{\tbox{PE}}\approx 0.386$~\cite{ABGR13} and
$\langle r_{\mathbb{R}}\rangle_{\tbox{GOE}}\approx 0.531$~\cite{ABGR13}.
Horizontal dashed lines in (b) correspond to the PE and GOE predictions for 
$\left\langle r_\mathbb{C}(\mathbf{A_{\tbox{H}}}) \right\rangle$:
$\langle r_{\mathbb{C}}\rangle_{\tbox{PE}}\approx 0.5~\cite{SRP20}$ and
$\langle r_{\mathbb{C}}\rangle_{\tbox{GOE}}\approx 0.569$~\cite{PRRCM20}.
Vertical dashed lines in (c,d) indicate the values of $\left\langle k \right\rangle$
chosen for the probability density functions reported in Fig.~\ref{Fig02}.
All averages are computed over $10^6$ ratios.}
\label{Fig01}
\end{figure}
\begin{figure*}[!t]
\includegraphics[width=0.8\textwidth]{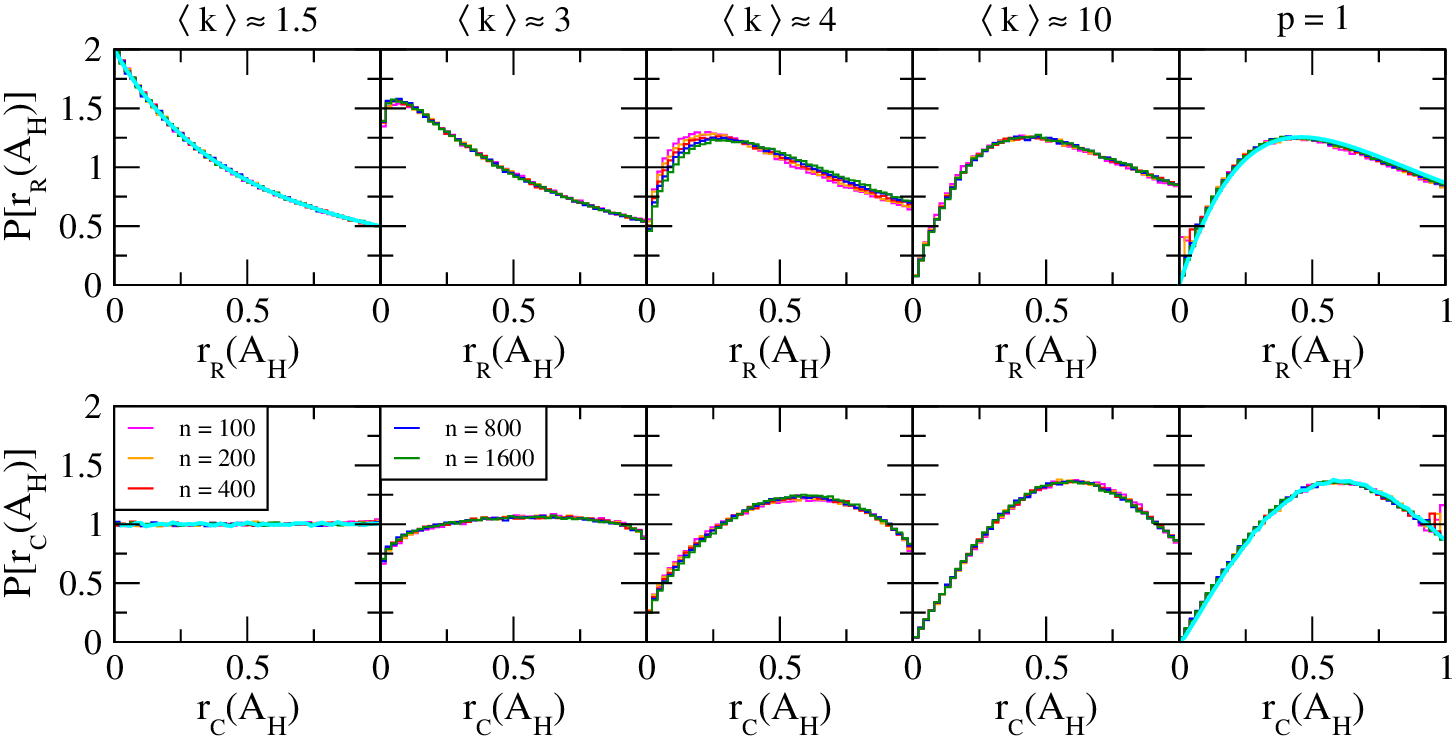}
\caption{Probability density function of the ratios 
$r_\mathbb{R}(\mathbf{A_{\tbox{H}}})$ (upper panels) and
$r_\mathbb{C}(\mathbf{A_{\tbox{H}}})$ (lower panels)
of undirected fuzzy Erd\"os-R\'enyi random graphs of size $n$.
Each histogram was constructed from the ratios of $10^6$ random graphs.
The average degree $\langle k \rangle$ is fixed in each column [see vertical dashed lines in Figs.~\ref{Fig01}(c,d)], 
except in the right panels where $p=1$ is used.
Solid cyan lines in upper-left and upper-right panels are Eqs.~(\ref{PrRPE}) and (\ref{PrRGOE}), respectively.
The solid cyan line in lower-left panel is Eq.~(\ref{PrCPE}).
The solid cyan line in lower-right panel was computed numerically from matrices of the GOE.}
\label{Fig02}
\end{figure*}

In Figs.~\ref{Fig01}(a) and~\ref{Fig01}(b) we present the average value 
of the eigenvalue spacing ratio $\left\langle r_\mathbb{R}(\mathbf{A_{\tbox{H}}}) \right\rangle$ and the 
average value of the ratio between nearest- and next-to-nearest neighbor eigenvalues
$\left\langle r_\mathbb{C}(\mathbf{A_{\tbox{H}}}) \right\rangle$ of 
undirected fuzzy ER random graphs as a function of the probability $p$.
We used graphs of five sizes: $n=100$, 200, 400, 800, and 1600.

From Figs.~\ref{Fig01}(a,b) we can see that
both ratios, $\left\langle r_\mathbb{R}(\mathbf{A_{\tbox{H}}}) \right\rangle$ and
$\left\langle r_\mathbb{C}(\mathbf{A_{\tbox{H}}}) \right\rangle$, transit from a minimum
to a maximum value as a function of $p$. 
Remarkably, the minimum and maximum values of both ratios coincide with the 
corresponding predictions for the Poisson Ensemble (i.e.~diagonal matrices with gaussian 
random entries) and the GOE.
We stress that this is remarkable because in the limits $p=0$ and $p=1$ the diagonal
random matrices and the full random matrices obtained from $\mathbf{A_{\tbox{H}}}$,
respectively, are not members of the Poisson Ensemble (PE) neither of the GOE.
We just observe a small-size effect in the curves of 
$\left\langle r_\mathbb{R}(\mathbf{A_{\tbox{H}}}) \right\rangle$ and
$\left\langle r_\mathbb{C}(\mathbf{A_{\tbox{H}}}) \right\rangle$ for $n=100$.

From Figs.~\ref{Fig01}(a,b) we can also observe that the curves 
$\left\langle r_\mathbb{R}(\mathbf{A_{\tbox{H}}}) \right\rangle$ vs.~$p$ 
and $\left\langle r_\mathbb{C}(\mathbf{A_{\tbox{H}}}) \right\rangle$ vs.~$p$ have 
the same shape for different values of $n$ but are displaced to the left on the $p-$axis
when increasing $n$. This indicates the existence of a parameter, dependent
on $n$, that may scale the curves of $\left\langle r_\mathbb{R}(\mathbf{A_{\tbox{H}}}) \right\rangle$ 
and $\left\langle r_\mathbb{C}(\mathbf{A_{\tbox{H}}}) \right\rangle$; see e.g.~\cite{AMRS20,PRRCM20,MA24}.
In fact, in previous papers we have reported that the average degree $\left\langle k \right\rangle$
serves as the scaling parameter of several topological and espectral properties of random graphs;
see e.g.~\cite{AMRS20,MAM26,MM26}.

Indeed, when plotting $\left\langle r_\mathbb{R}(\mathbf{A_{\tbox{H}}}) \right\rangle$ vs.~$\left\langle k \right\rangle$ 
and $\left\langle r_\mathbb{C}(\mathbf{A_{\tbox{H}}}) \right\rangle$ vs.~$\left\langle k \right\rangle$,
with~\cite{MM26}
\begin{equation}
\left\langle k \right\rangle = 1 + (n-1)p ,
\label{k}
\end{equation}
we observe that the curves for different $n$ collapse on top of a universal curve; see Figs.~\ref{Fig01}(c,d).
Here, the small-size effect of the curves corresponding to $n=100$ is quite evident: They are close but do not 
fall on top of the curves with larger $n$.

From Figs.~\ref{Fig01}(c,d) it is clear that $\left\langle k \right\rangle$ scales both ratios 
$\left\langle r_\mathbb{R}(\mathbf{A_{\tbox{H}}}) \right\rangle$ 
and $\left\langle r_\mathbb{C}(\mathbf{A_{\tbox{H}}}) \right\rangle$.
Moreover, $\left\langle k \right\rangle$ is expected to also scale the corresponding probability density functions
(PDFs), see e.g.~\cite{PRRCM20,MA24}.
This is verified in Fig.~\ref{Fig02} where we plot $P[r_\mathbb{R}(\mathbf{A_{\tbox{H}}})]$ (upper panels) and
$P[r_\mathbb{C}(\mathbf{A_{\tbox{H}}})]$ (lower panels) of undirected fuzzy ER random graphs.
Each panel (except the right ones) corresponds to a fixed average degree $\left\langle k \right\rangle$.
Specifically, we choose four values of $\left\langle k \right\rangle$ along the transition from isolated nodes 
to complete graphs, as indicated by the horizontal dashed lines in 
Figs.~\ref{Fig01}(c,d): $\left\langle k \right\rangle=1.5$, 3, 4, and 10. In addition, the right panels of Fig.~\ref{Fig02}
correspond to $p=1$ (complete graphs).
Note that each panel in Fig.~\ref{Fig02} contains five histograms corresponding to graphs of different sizes.
As expected, once the average degree $\left\langle k \right\rangle$ is fixed, neither
$P[r_\mathbb{R}(\mathbf{A_{\tbox{H}}})]$ nor $P[r_\mathbb{C}(\mathbf{A_{\tbox{H}}})]$
depend on the graph size (we only see small-size effects for $P[r_\mathbb{R}(\mathbf{A_{\tbox{H}}})]$
at intermediate values of $\left\langle k \right\rangle$, see the upper-middle panel in Fig.~\ref{Fig02}).

Also, note that when $\left\langle k \right\rangle = 1$, $P[r_\mathbb{R}(\mathbf{A_{\tbox{H}}})]$ and 
$P[r_\mathbb{C}(\mathbf{A_{\tbox{H}}})]$ are well reproduced by the RMT predictions for the
PE~\cite{ABGR13} 
\begin{equation}
P_{\tbox{PE}}(r_\mathbb{R}) = \frac{2}{(1+r_\mathbb{R})^2} 
\label{PrRPE}
\end{equation}
and~\cite{SRP20}
\begin{equation}
P_{\tbox{PE}}(r_\mathbb{C}) = 1 ,
\label{PrCPE}
\end{equation}
see the cyan curves in the left panels.
While for $p=1$, $P[r_\mathbb{R}(\mathbf{A_{\tbox{H}}})]$ and $P[r_\mathbb{C}(\mathbf{A_{\tbox{H}}})]$
correspond to the RMT prediction for the GOE, see the cyan curves in the right panels.
However, for the GOE, only $P_{\tbox{GOE}}(r_\mathbb{R})$ is known analytically~\cite{ABGR13}:
\begin{equation}
P_{\tbox{GOE}}(r_\mathbb{R}) = 
\frac{27}{4} \frac{r_\mathbb{R}+r_\mathbb{R}^2}{(1+r_\mathbb{R}+r_\mathbb{R}^2)^{5/2}} .
\label{PrRGOE}
\end{equation}
So, the cyan curve in the lower-right panel corresponding to $P_{\tbox{GOE}}(r_\mathbb{C})$ was
computed numerically from matrices of the GOE.

\begin{figure}[t]
\centering
\includegraphics[width=\columnwidth]{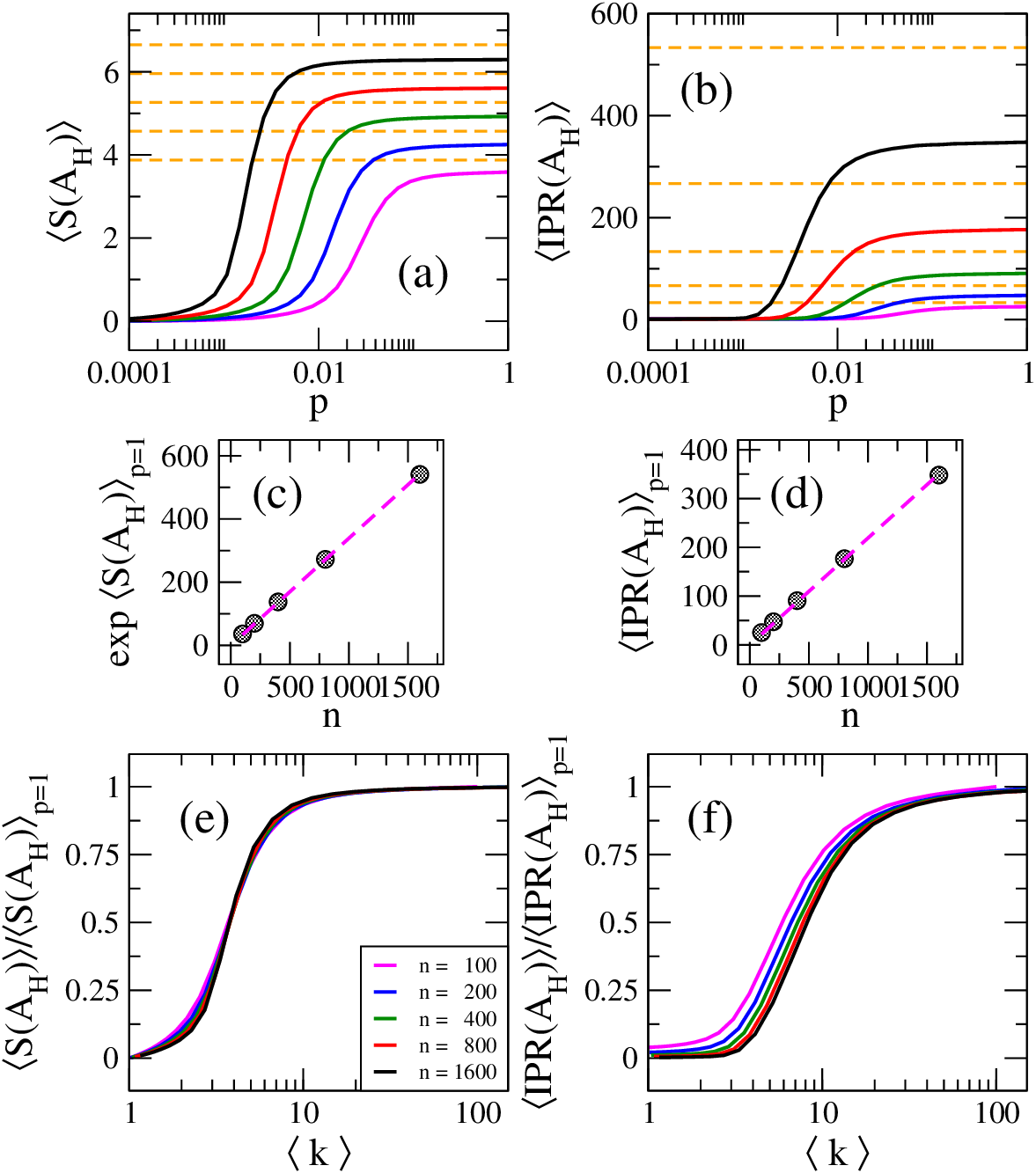}
\caption{
(a) Average Shannon entropy $\left\langle S(\mathbf{A_{\tbox{H}}}) \right\rangle$ and 
(b) average inverse participation ratio $\left\langle \mbox{IPR(}\mathbf{A_{\tbox{H}}}) \right\rangle$ of 
undirected fuzzy Erd\"os-R\'enyi random graphs (of size $n$) 
as a function of the probability $p$.
Horizontal dashed lines in (a) and (b) correspond to the GOE predictions for 
$\left\langle S(\mathbf{A_{\tbox{H}}}) \right\rangle$ and 
$\left\langle \mbox{IPR(}\mathbf{A_{\tbox{H}}}) \right\rangle$:
$\langle S \rangle_{\tbox{GOE}}\approx \ln(n/2.07)$ and
$\langle \mbox{IPR}\rangle_{\tbox{GOE}}\approx n/3$.
(c) $\exp[\left\langle S(\mathbf{A_{\tbox{H}}}) \right\rangle]$ at $p=1$ as a function of $n$.
The dashed line is a fitting of the data with the function 
$\exp[\left\langle S(\mathbf{A_{\tbox{H}}}) \right\rangle]=n/x$ with $x\approx 2.95$.
(d) $\left\langle \mbox{IPR}(\mathbf{A_{\tbox{H}}}) \right\rangle$ at $p=1$ as a function of $n$.
The dashed line is a fitting of the data with the function 
$\left\langle \mbox{IPR}(\mathbf{A_{\tbox{H}}}) \right\rangle=n/x$ with $x\approx 4.57$.
(e) $\left\langle S(\mathbf{A_{\tbox{H}}}) \right\rangle/\left\langle S(\mathbf{A_{\tbox{H}}}) \right\rangle_{p=1}$ and 
(f) $\left\langle \mbox{IPR}(\mathbf{A_{\tbox{H}}}) \right\rangle/\left\langle \mbox{IPR}(\mathbf{A_{\tbox{H}}}) \right\rangle_{p=1}$
as a function of the average degree $\left\langle k \right\rangle$.
All averages are computed over $10^6/n$ random graphs.}
\label{Fig03}
\end{figure}

Now we turn to eigenvector properties.
Then in Figs.~\ref{Fig03}(a) and~\ref{Fig03}(b) we present the average Shannon entropy 
$\left\langle S(\mathbf{A_{\tbox{H}}}) \right\rangle$ and the average inverse participation ratio 
$\left\langle \mbox{IPR(}\mathbf{A_{\tbox{H}}}) \right\rangle$ of 
undirected fuzzy ER random graphs as a function of the probability $p$.
We again use graphs of five sizes: $n=100$, 200, 400, 800, and 1600.

As well as for the ratios, $\left\langle r_\mathbb{R}(\mathbf{A_{\tbox{H}}}) \right\rangle$ and
$\left\langle r_\mathbb{C}(\mathbf{A_{\tbox{H}}}) \right\rangle$, here, 
$\left\langle S(\mathbf{A_{\tbox{H}}}) \right\rangle$ and  
$\left\langle \mbox{IPR(}\mathbf{A_{\tbox{H}}}) \right\rangle$ transit from a minimum
to a maximum value as a function of $p$. 
The minimum values of $\left\langle S(\mathbf{A_{\tbox{H}}}) \right\rangle$ and  
$\left\langle \mbox{IPR(}\mathbf{A_{\tbox{H}}}) \right\rangle$, as expected, correspond
to those of diagonal matrices: 0 and 1, respectively.
However, in contrast with $\left\langle r_\mathbb{R}(\mathbf{A_{\tbox{H}}}) \right\rangle$ and
$\left\langle r_\mathbb{C}(\mathbf{A_{\tbox{H}}}) \right\rangle$, the maximum values of
$\left\langle S(\mathbf{A_{\tbox{H}}}) \right\rangle$ and  
$\left\langle \mbox{IPR(}\mathbf{A_{\tbox{H}}}) \right\rangle$ do not correspond to the predictions
for the GOE, which read as $\langle S \rangle_{\tbox{GOE}}\approx \ln(n/2.07)$ and
$\langle \mbox{IPR}\rangle_{\tbox{GOE}}\approx n/3$; see the horizontal dashed lines 
in Figs.~\ref{Fig03}(a) and~\ref{Fig03}(b).
Instead, we found
\begin{equation}
\left\langle S(\mathbf{A_{\tbox{H}}}) \right\rangle_{p=1} \approx \ln\left( \frac{n}{2.95} \right)
\label{SA}
\end{equation}
and 
\begin{equation}
\left\langle \mbox{IPR}(\mathbf{A_{\tbox{H}}}) \right\rangle_{p=1} \approx \frac{n}{4.57} .
\label{IPRA}
\end{equation}
We obtained the dependencies above by plotting $\exp[\left\langle S(\mathbf{A_{\tbox{H}}}) \right\rangle_{p=1}]$
vs.~$n$ and $\left\langle \mbox{IPR}(\mathbf{A_{\tbox{H}}}) \right\rangle_{p=1}$ vs.~$n$ and 
performing fittings of the form $\exp[\left\langle S(\mathbf{A_{\tbox{H}}}) \right\rangle]=n/x$ 
and $\left\langle \mbox{IPR}(\mathbf{A_{\tbox{H}}}) \right\rangle=n/x$, respectively, with $x$ as the
fitting parameter; see Figs.~\ref{Fig03}(c) and~\ref{Fig03}(d).

Then, in Figs.~\ref{Fig03}(e) and~\ref{Fig03}(f) we plot, respectively,
$\left\langle S(\mathbf{A_{\tbox{H}}}) \right\rangle/\left\langle S(\mathbf{A_{\tbox{H}}}) \right\rangle_{p=1}$ and 
$\left\langle \mbox{IPR}(\mathbf{A_{\tbox{H}}}) \right\rangle/\left\langle \mbox{IPR}(\mathbf{A_{\tbox{H}}}) \right\rangle_{p=1}$
as a function of the average degree and verify that 
$\left\langle k \right\rangle$ serves as the scaling parameter of the normalized Shannon entropies and inverse 
participation ratios of the eigenvectors of $\mathbf{A_{\tbox{H}}}$.
We just want to note that a small-size effect is clearly observed for 
$\left\langle \mbox{IPR}(\mathbf{A_{\tbox{H}}}) \right\rangle/\left\langle \mbox{IPR}(\mathbf{A_{\tbox{H}}}) \right\rangle_{p=1}$,
see Fig.~\ref{Fig03}(f).

\subsection{Directed fuzzy Erd\"os-R\'enyi random graphs}

Recently, the singular-value statistics (SVS) has been presented as a RMT tool able to properly 
characterize non-Hermitian random matrix ensembles~\cite{KXOS23,HTM24} as well as to identify
the delocalization transition in non-Hermitian many-body systems~\cite{RBSC24} and models of 
directed networks~\cite{MA24,THM25}. So, we also use SVS here to characterize the
spectral properties of $\mathbf{A_{\tbox{nH}}}$ as follows: 
Given the ordered square roots of the real eigenvalues of 
the Hermitian matrix $\mathbf{A_{\tbox{nH}}}\mathbf{A_{\tbox{nH}}}^\dagger$, $s_1>s_2>\cdots >s_N$ 
(which are the singular values of $\mathbf{A_{\tbox{nH}}}$), we compute the ratio 
$r_\mathbb{R}(\mathbf{A_{\tbox{nH}}}\mathbf{A_{\tbox{nH}}}^\dagger)$
between consecutive singular-value spacings, 
where the $i$--th ratio is given by~\cite{KXOS23}
\begin{equation}
r^i_\mathbb{R}(\mathbf{A_{\tbox{nH}}}\mathbf{A_{\tbox{nH}}}^\dagger) = \frac{\mathrm{min}(s_{i+1}-s_i,s_{i}-s_{i-1})}{\mathrm{max}(s_{i+1}-s_i,s_i-s_{i-1})} .
\label{rSV}
\end{equation}
Above, as usual, $\mathbf{A_{\tbox{nH}}}^\dagger$ is the conjugate transpose of $\mathbf{A_{\tbox{nH}}}$.
Moreover, since for real matrices, as the ones we consider here, the conjugate transpose is just the 
transpose $\mathbf{A_{\tbox{nH}}}^\dagger=\mathbf{A_{\tbox{nH}}}^{\tbox{T}}$,
then, in what follows, the SVS concerns the spectra of $\mathbf{A_{\tbox{nH}}}\mathbf{A_{\tbox{nH}}}^{\tbox{T}}$
and $r_\mathbb{R}(\mathbf{A_{\tbox{nH}}}\mathbf{A_{\tbox{nH}}}^\dagger) \equiv r_\mathbb{R}(\mathbf{A_{\tbox{nH}}}\mathbf{A_{\tbox{nH}}}^{\tbox{T}})$.

Below we use exact numerical diagonalization to obtain the complex eigenvalues $\lambda_i$, the singular 
values $s_i$, and right eigenvectors $\Psi^i$ ($i =1,\ldots,n$) of ensembles of non-Hermitian adjacency matrices 
$\mathbf{A_{\tbox{nH}}}$ characterized by the parameter pair $(n,p)$.

\begin{figure}[t]
\centering
\includegraphics[width=0.95\columnwidth]{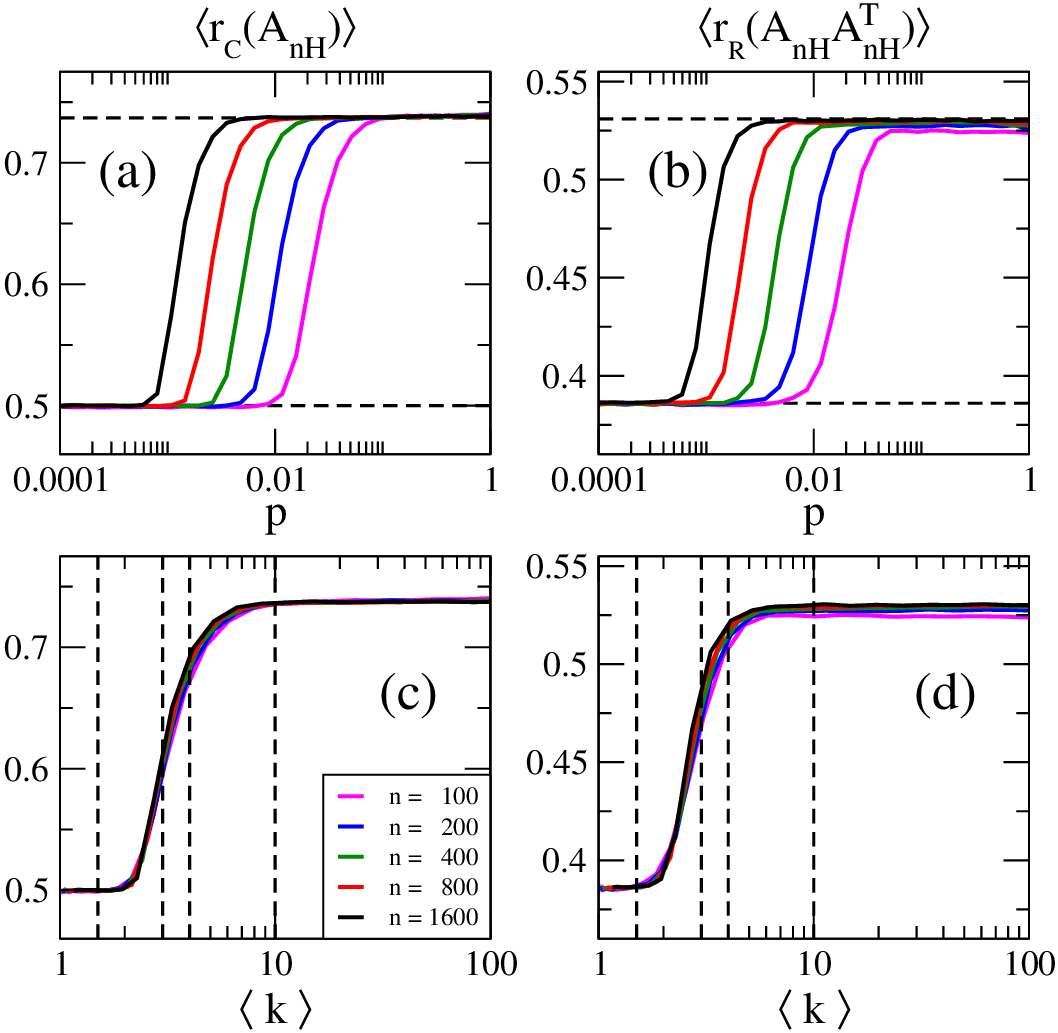}
\caption{Average ratios 
(a,c) $\left\langle r_\mathbb{C}(\mathbf{A_{\tbox{nH}}}) \right\rangle$ and 
(b,d) $\left\langle r_\mathbb{R}(\mathbf{A_{\tbox{nH}}A_{\tbox{nH}}^{\tbox{T}}}) \right\rangle$ of 
directed fuzzy Erd\"os-R\'enyi random graphs (of size $n$) 
as a function of (a,b) the probability $p$ and (c,d) the average degree $\left\langle k \right\rangle$.
Horizontal dashed lines in (a) correspond to the PE and GinOE predictions for 
$\left\langle r_\mathbb{C}(\mathbf{A_{\tbox{nH}}}) \right\rangle$:
$\langle r_{\mathbb{C}}\rangle_{\tbox{PE}}\approx 0.5$~\cite{SRP20} and
$\langle r_{\mathbb{C}}\rangle_{\tbox{GinOE}}\approx 0.737$~\cite{SRP20}.
Horizontal dashed lines in (b) correspond to the PE and GOE predictions for 
$\left\langle r_\mathbb{R}(\mathbf{A_{\tbox{nH}}A_{\tbox{nH}}^{\tbox{T}}}) \right\rangle$:
$\langle r_{\mathbb{R}}\rangle_{\tbox{PE}}\approx 0.386$~\cite{ABGR13} and
$\langle r_{\mathbb{R}}\rangle_{\tbox{GOE}}\approx 0.531$~\cite{ABGR13}.
Vertical dashed lines in (c,d) indicate the values of $\left\langle k \right\rangle$
chosen for the probability density functions reported in Fig.~\ref{Fig05}.
All averages are computed over $10^6$ ratios.}
\label{Fig04}
\end{figure}

\begin{figure*}[!t]
\includegraphics[width=0.8\textwidth]{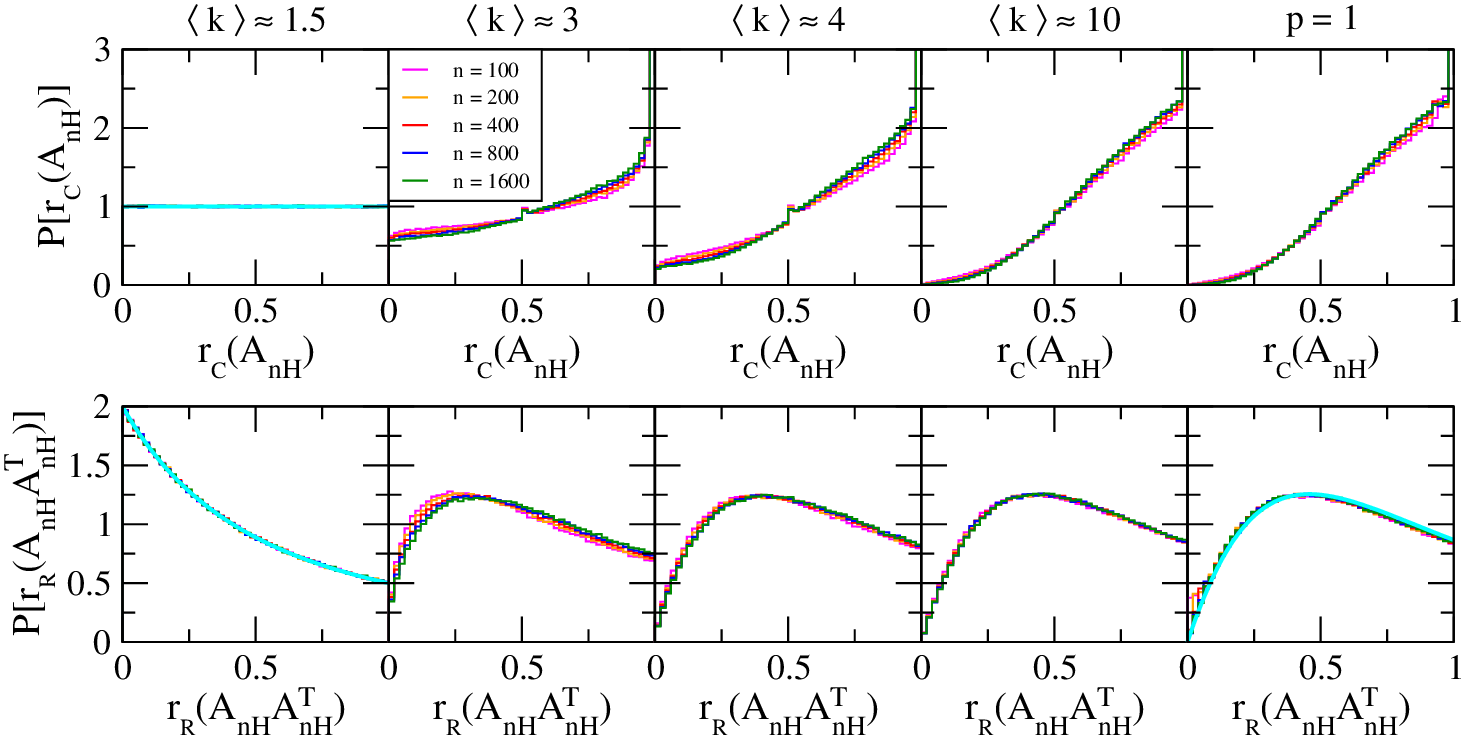}
\caption{Probability density function of the ratios 
$r_\mathbb{C}(\mathbf{A_{\tbox{nH}}})$ (upper panels) and
$r_\mathbb{R}(\mathbf{A_{\tbox{nH}}A_{\tbox{nH}}^{\tbox{T}}})$ (lower panels)
of directed fuzzy Erd\"os-R\'enyi random graphs of size $n$.
Each histogram was constructed from the ratios of $10^6$ random graphs.
The average degree $\langle k \rangle$ is fixed in each column [see vertical dashed lines in Figs.~\ref{Fig01}(c,d)], 
except in the right panels where $p=1$ is used.
The solid cyan line in upper-left panel is Eq.~(\ref{PrCPE}).
Solid cyan lines in lower-left and lower-right panels are Eqs.~(\ref{PrRPE}) and (\ref{PrRGOE}), respectively.}
\label{Fig05}
\end{figure*}

\begin{figure}[t]
\centering
\includegraphics[width=\columnwidth]{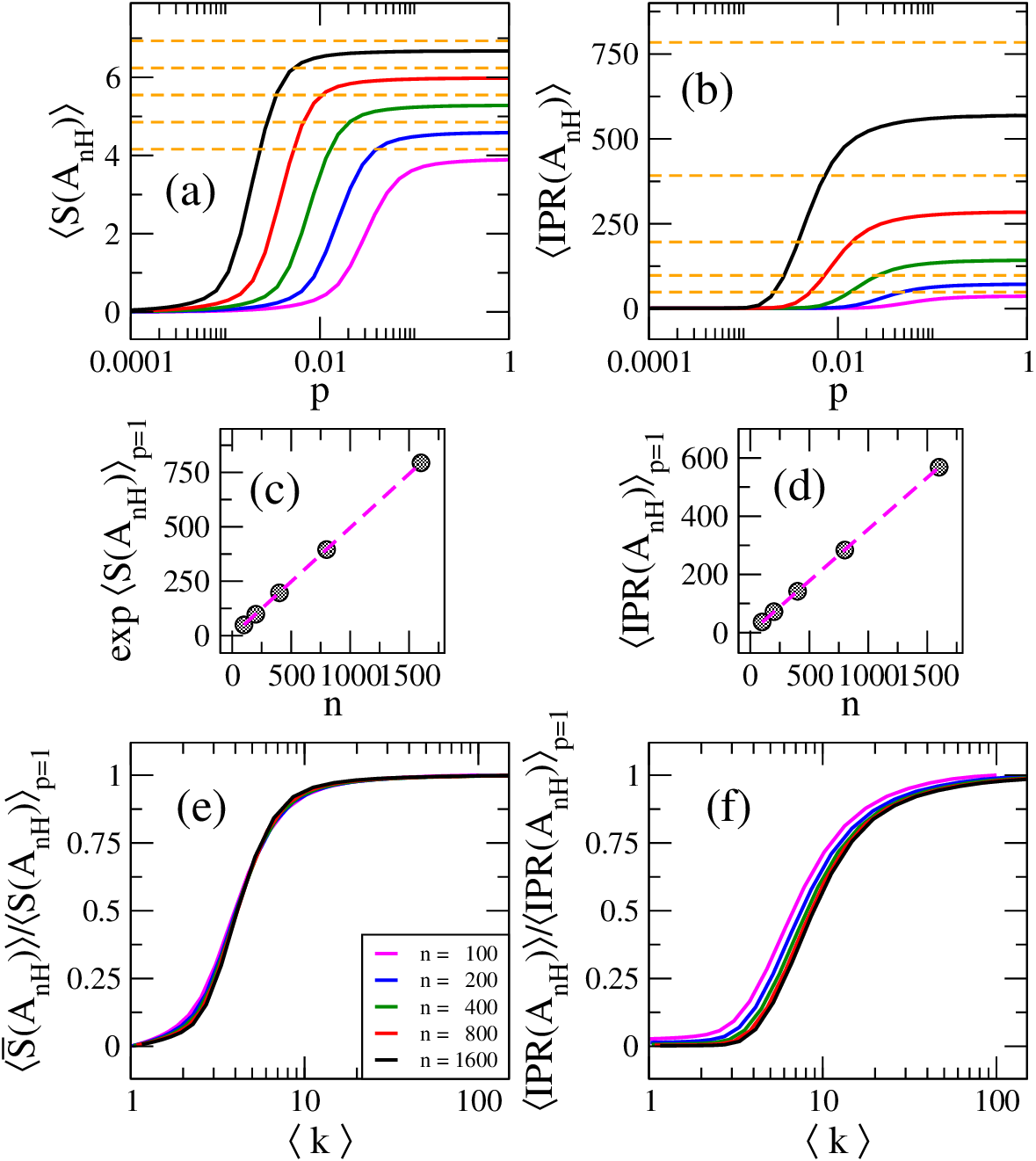}
\caption{
(a) Average Shannon entropy $\left\langle S(\mathbf{A_{\tbox{nH}}}) \right\rangle$ and 
(b) average inverse participation ratio $\left\langle \mbox{IPR(}\mathbf{A_{\tbox{nH}}}) \right\rangle$ of 
directed fuzzy Erd\"os-R\'enyi random graphs (of size $n$) 
as a function of the probability $p$.
Horizontal dashed lines in (a) and (b) correspond to the GinOE predictions for 
$\left\langle S(\mathbf{A_{\tbox{nH}}}) \right\rangle$ and 
$\left\langle \mbox{IPR(}\mathbf{A_{\tbox{nH}}}) \right\rangle$:
$\langle S \rangle_{\tbox{GinOE}}\approx \ln(n/1.56)$~\cite{PRRCM20} and
$\langle \mbox{IPR}\rangle_{\tbox{GinOE}}\approx n/2.04$~\cite{PM23}.
(c) $\exp[\left\langle S(\mathbf{A_{\tbox{nH}}}) \right\rangle]$ at $p=1$ as a function of $n$.
The dashed line is a fitting of the data with the function 
$\exp[\left\langle S(\mathbf{A_{\tbox{nH}}}) \right\rangle]=n/x$ with $x\approx 2$.
(d) $\left\langle \mbox{IPR}(\mathbf{A_{\tbox{nH}}}) \right\rangle$ at $p=1$ as a function of $n$.
The dashed line is a fitting of the data with the function 
$\left\langle \mbox{IPR}(\mathbf{A_{\tbox{nH}}}) \right\rangle=n/x$ with $x\approx 2.8$.
(e) $\left\langle S(\mathbf{A_{\tbox{nH}}}) \right\rangle/\left\langle S(\mathbf{A_{\tbox{nH}}}) \right\rangle_{p=1}$ and 
(f) $\left\langle \mbox{IPR}(\mathbf{A_{\tbox{nH}}}) \right\rangle/\left\langle \mbox{IPR}(\mathbf{A_{\tbox{nH}}}) \right\rangle_{p=1}$
as a function of the average degree $\left\langle k \right\rangle$.
All averages are computed over $10^6/n$ random graphs.}
\label{Fig06}
\end{figure}

Taking as a reference the results reported in Figs.~\ref{Fig01}-\ref{Fig03} for undirected fuzzy ER 
random graphs, in Figs.~\ref{Fig04}-\ref{Fig06} we present 
the average value of the ratio between nearest- and next-to-nearest neighbor eigenvalues 
$\left\langle r_\mathbb{C}(\mathbf{A_{\tbox{nH}}}) \right\rangle$ and the average value of the 
eigenvalue spacing ratio $\left\langle r_\mathbb{R}(\mathbf{A_{\tbox{nH}}A_{\tbox{nH}}^{\tbox{T}}}) \right\rangle$,
the corresponding PDFs, $P[r_\mathbb{C}(\mathbf{A_{\tbox{nH}}})]$ and 
$P[r_\mathbb{R}(\mathbf{A_{\tbox{nH}}A_{\tbox{nH}}^{\tbox{T}}})]$, and
the verage Shannon entropy $\left\langle S(\mathbf{A_{\tbox{nH}}}) \right\rangle$ and 
the average inverse participation ratio $\left\langle \mbox{IPR(}\mathbf{A_{\tbox{nH}}}) \right\rangle$ 
of directed fuzzy Erd\"os-R\'enyi random graphs.
For comparison purposes, Figs.~\ref{Fig04}-\ref{Fig06} have the same format and color codes as Figs.~\ref{Fig01}-\ref{Fig03}.

Indeed, from Figs.~\ref{Fig04}-\ref{Fig06} we can make similar observations as those made from 
Figs.~\ref{Fig01}-\ref{Fig03} in the previous Subsection.
Namely:
\begin{itemize}
\item[(i)]
Both ratios, $\left\langle r_\mathbb{C}(\mathbf{A_{\tbox{nH}}}) \right\rangle$ and 
$\left\langle r_\mathbb{R}(\mathbf{A_{\tbox{nH}}A_{\tbox{nH}}^{\tbox{T}}}) \right\rangle$, transit from a minimum
to a maximum value as a function of $p$; see Figs.~\ref{Fig04}(a,b).

\item[(ii)]
Remarkably, the minimum and maximum values of both ratios coincide with the 
corresponding predictions for the PE and the GinOE; see the dashed lines in Figs.~\ref{Fig04}(a,b).
We stress that this is remarkable because in the limits $p=0$ and $p=1$ the diagonal
random matrices and the full random matrices obtained from $\mathbf{A_{\tbox{nH}}}$,
respectively, are not members of the PE neither of the GinOE.

\item[(iii)]
The curves $\left\langle r_\mathbb{C}(\mathbf{A_{\tbox{nH}}}) \right\rangle$ vs.~$p$ 
and $\left\langle r_\mathbb{R}(\mathbf{A_{\tbox{nH}}A_{\tbox{nH}}^{\tbox{T}}}) \right\rangle$ vs.~$p$ have 
the same shape for different values of $n$ but are displaced to the left on the $p-$axis
when increasing $n$; see Figs.~\ref{Fig04}(a,b).

\item[(iv)]
When plotting $\left\langle r_\mathbb{C}(\mathbf{A_{\tbox{nH}}}) \right\rangle$ vs.~$\left\langle k \right\rangle$ 
and $\left\langle r_\mathbb{R}(\mathbf{A_{\tbox{nH}}A_{\tbox{nH}}^{\tbox{T}}}) \right\rangle$ vs.~$\left\langle k \right\rangle$,
the curves for different $n$ collapse on top of a universal curve; see Figs.~\ref{Fig04}(c,d).
That is, $\left\langle k \right\rangle$ scales both ratios 
$\left\langle r_\mathbb{C}(\mathbf{A_{\tbox{nH}}}) \right\rangle$ and 
$\left\langle r_\mathbb{R}(\mathbf{A_{\tbox{nH}}A_{\tbox{nH}}^{\tbox{T}}}) \right\rangle$.

\item[(v)]
We observe a small-size effect in the curves of 
$\left\langle r_\mathbb{C}(\mathbf{A_{\tbox{nH}}}) \right\rangle$ and 
$\left\langle r_\mathbb{R}(\mathbf{A_{\tbox{nH}}A_{\tbox{nH}}^{\tbox{T}}}) \right\rangle$ for $n=100$; 
see the magenta curves in Fig.~\ref{Fig04}.

\item[(vi)]
$\left\langle k \right\rangle$ also scales the PDFs of $r_\mathbb{C}(\mathbf{A_{\tbox{nH}}})$ and 
$r_\mathbb{R}(\mathbf{A_{\tbox{nH}}A_{\tbox{nH}}^{\tbox{T}}})$.
That is, once $\left\langle k \right\rangle$ is fixed, neither
$P[r_\mathbb{C}(\mathbf{A_{\tbox{nH}}})]$ nor $P[r_\mathbb{R}(\mathbf{A_{\tbox{nH}}A_{\tbox{nH}}^{\tbox{T}}})]$
depend on the graph size $n$; see Fig.~\ref{Fig05}. 

\item[(vii)]
We observe small-size effects in $P[r_\mathbb{C}(\mathbf{A_{\tbox{nH}}})]$ for any $\left\langle k \right\rangle>1$ 
and in $P[r_\mathbb{R}(\mathbf{A_{\tbox{nH}}A_{\tbox{nH}}^{\tbox{T}}})]$ at intermediate values of 
$\left\langle k \right\rangle$; see Fig.~\ref{Fig05}.

\item[(viii)]
When $\left\langle k \right\rangle = 1$, $P[r_\mathbb{C}(\mathbf{A_{\tbox{nH}}})]$ and 
$P[r_\mathbb{R}(\mathbf{A_{\tbox{nH}}A_{\tbox{nH}}^{\tbox{T}}})]$ are well reproduced by the RMT predictions 
for the PE; see the cyan curves in the left panels of Fig.~\ref{Fig05}.

\item[(ix)]
When $p = 1$, $P[r_\mathbb{R}(\mathbf{A_{\tbox{nH}}A_{\tbox{nH}}^{\tbox{T}}})]$ is 
well reproduced by the RMT prediction 
for the GinOE; see the cyan curve in the lower-right panel of Fig.~\ref{Fig05}.

\item[(x)]
$\left\langle S(\mathbf{A_{\tbox{nH}}}) \right\rangle$ and  
$\left\langle \mbox{IPR(}\mathbf{A_{\tbox{nH}}}) \right\rangle$ transit from a minimum
to a maximum value as a function of $p$; see Figs.~\ref{Fig06}(a,b).

\item[(xi)]
The minimum values of $\left\langle S(\mathbf{A_{\tbox{nH}}}) \right\rangle$ and  
$\left\langle \mbox{IPR(}\mathbf{A_{\tbox{nH}}}) \right\rangle$, as expected, correspond
to those of diagonal matrices: 0 and 1, respectively; see Figs.~\ref{Fig06}(a,b).

\item[(xii)]
In contrast with $\left\langle r_\mathbb{C}(\mathbf{A_{\tbox{nH}}}) \right\rangle$ and 
$\left\langle r_\mathbb{R}(\mathbf{A_{\tbox{nH}}A_{\tbox{nH}}^{\tbox{T}}}) \right\rangle$, 
the maximum values of $\left\langle S(\mathbf{A_{\tbox{nH}}}) \right\rangle$ and  
$\left\langle \mbox{IPR(}\mathbf{A_{\tbox{nH}}}) \right\rangle$ do not correspond to the
predictions for the GinOE; see the horizontal dashed lines in Figs.~\ref{Fig06}(a,b).
Instead, we found
\begin{equation}
\left\langle S(\mathbf{A_{\tbox{nH}}}) \right\rangle_{p=1} \approx \ln\left( \frac{n}{2} \right)
\label{SnA}
\end{equation}
and 
\begin{equation}
\left\langle \mbox{IPR}(\mathbf{A_{\tbox{nH}}}) \right\rangle_{p=1} \approx \frac{n}{2.8} ;
\label{IPRnA}
\end{equation}
see Figs.~\ref{Fig06}(c,d),

\item[(xiii)]
$\left\langle k \right\rangle$ serves as the scaling parameter of the normalized Shannon entropies and inverse 
participation ratios of the eigenvectors of $\mathbf{A_{\tbox{nH}}}$; Figs.~\ref{Fig06}(d,f).

\item[(xiv)]
We observe a clear small-size effect for
$\left\langle \mbox{IPR}(\mathbf{A_{\tbox{nH}}}) \right\rangle/\left\langle \mbox{IPR}(\mathbf{A_{\tbox{nH}}}) \right\rangle_{p=1}$, 
see Fig.~\ref{Fig06}(f).

\end{itemize}

\section{Discussion and conclusions}

By considering the membership values of the vertices and edges of a fuzzy ER graph model 
as random variables, the corresponding adjacency matrices produced two random matrix ensembles:
an Hermitian ensemble $\{\mathbf{A_{\tbox{H}}}\}$ and a non-Hermitian ensemble 
$\{\mathbf{A_{\tbox{nH}}}\}$ representing, respectively, undirected and directed fuzzy graphs. 
Notice that both ensembles transit form diagonal random matrices to full random matrices by increasing the
connection probability $p$, of the underlying ER model, from 0 to 1 for any given graph size $n$.

Since none of these two random matrix ensembles was studied before, in this work, we performed a 
detailed numerical study of the spectral statistics of $\{\mathbf{A_{\tbox{H}}}\}$ and $\{\mathbf{A_{\tbox{nH}}}\}$ 
by the use of standard RMT measures:
The average value of the eigenvalue spacing ratio $r_\mathbb{R}$ and the average value of 
the ratio between nearest- and next-to-nearest neighbor eigenvalues $r_\mathbb{C}$; as well as
the average value of the Shannon entropies $S$ and the inverse participation ratios $\mbox{IPR}$ 
of the eigenvectors.

We showed that the average degree of the ER model, see Eq.~(\ref{k}), serves as the scaling parameter of the 
spectral properties of the corresponding fuzzy setups; see Figs.~\ref{Fig01}(c,d), \ref{Fig02}, \ref{Fig03}(e,f),
\ref{Fig04}(c,d), \ref{Fig05}, and~\ref{Fig06}(e,f).

Moreover, remarkably, when $p=0$ and $p=1$ the eigenvalue properties of $\{\mathbf{A_{\tbox{H}}}\}$
[$\{\mathbf{A_{\tbox{nH}}}\}$] coincide with those of the PE and the GOE [GinOE], respectively.
However, when $p=1$, the eigenvector properties of both $\{\mathbf{A_{\tbox{H}}}\}$ and
$\{\mathbf{A_{\tbox{nH}}}\}$ do not correspond to those for the GOE and the GinOE, respectively. 
Therefore, we obtained phenomenological expressions for 
$\left\langle S(\mathbf{A_{\tbox{H}}}) \right\rangle_{p=1}$,
$\left\langle \mbox{IPR}(\mathbf{A_{\tbox{H}}}) \right\rangle_{p=1}$,
$\left\langle S(\mathbf{A_{\tbox{nH}}}) \right\rangle_{p=1}$, and
$\left\langle \mbox{IPR}(\mathbf{A_{\tbox{nH}}}) \right\rangle_{p=1}$ which are given in
Eqs.~(\ref{SA}), (\ref{IPRA}), (\ref{SnA}), and~(\ref{IPRnA}), respectively.

We hope that this work may motivate further numerical and well as analytical studies of fuzzy random 
graphs from a statistical RMT perspective.

\section*{Acknowledgements}
We thank support from SECIHTI (Grant No.~CBF-2025-I-2236), Mexico.

%====================================================================++

\end{document}